\def\emph#1{{\rm#1}}
\def\sign{{\rm sign}}

\documentstyle{ioplppt}
\jl{1} % journal of physics A
\begin{document}
\input epsf
%\draft
%\preprint{T97/xxx}
\title{Coarsening and persistence in a class of stochastic processes interpolating between the Ising
and voter models}[Coarsening and persistence in a class of stochastic processes]

\author{J-M~Drouffe\dag\, and C~Godr\`eche\ddag \S}

\address{\dag\ Service de Physique Th\'eorique,
CEA-Saclay, 91191 Gif sur Yvette cedex, France}

\address{\ddag\ Service de Physique de l'\'Etat Condens\'e,
CEA-Saclay, 91191 Gif sur Yvette cedex, France}

\address{\S\ Laboratoire de Physique Th\'eorique et Mod\'elisation,
Universit\'e de Cergy Pontoise, France}

\begin{abstract}
We study the dynamics of a class of two dimensional stochastic processes, depending on
two parameters, which may be interpreted as two different temperatures, 
respectively associated to interfacial and to bulk noise.
Special lines in the plane of parameters correspond to the Ising model, voter
model and majority vote model.
The dynamics of this class of models may be described formally in terms of 
reaction diffusion processes for a set of 
coalescing, annihilating, and branching random walkers.
We use the freedom allowed by the space of parameters to measure, by numerical
simulations, the persistence probability of a generic model in the low temperature
phase, where the system coarsens.
This probability is found to decay at large times as a power law with 
a seemingly constant exponent
$\theta\approx 0.22$.
We also discuss the
connection between persistence and the nature of the interfaces between domains.
\end{abstract}

{\it submitted for publication to \JPA}


\maketitle

%-----------------------------------------------------------------------
\section{Introduction\label{intro}}
%-----------------------------------------------------------------------

As is well known, an Ising system quenched from high temperature to low
temperature exhibits coarsening in its temporal evolution 
\cite{langer,brayRevue}.
This property holds in any dimension.
The voter model \cite{ligg}, defined as a purely dynamical system, is identical
to the Ising model with Glauber \cite{glauber} or heat bath dynamics in one
dimension but its behaviour progressively departs from that of the latter when
dimension increases. 
For instance the two dimensional voter model also exhibits
properties similar to coarsening \cite{cox,spohn}, 
though the way the system coarsens differs, in
some respect, from that of the two dimensional Ising model. 
Measuring the persistence probability
of the two models demonstrates more strikingly the difference in behaviour
between them. 
While the persistence probability for the two dimensional Ising
model at zero temperature decays as
$t^{-\theta}$ with
$\theta\approx 0.22$ \cite{der1,stau1,der2}, it behaves as 
$\exp[-A\,\ln^2 t]$ for the two dimensional voter model \cite{ben,how}. 

In an attempt to understand the differences in behaviour between the dynamics of
the two dimensional Ising and voter models, we were naturally lead to
introduce a whole class of models interpolating between them. 
The idea stems from the observation that the
rules for updating a spin in the voter model resemble those used for the Ising
model, in the Glauber or heat bath algorithms. 
This class of models is defined in any dimension $d$, but the present study is
restricted to two dimensions, where the models depend continuously on two
parameters, which may be interpreted as two temperatures. 

The aim of the present work is to use the freedom allowed by the space of parameters
in order to investigate the mechanisms by which coarsening and persistence continuously
change when going from the Ising model to the voter model.
We will see that physical insight is provided by interpreting the two parameters
defining the dynamics as two different temperatures, respectively
associated to interfacial and to bulk noise.
We shall also investigate whether the persistence exponent $\theta$,
which seems constant $\approx 0.22$ for the two dimensional Ising model when temperature
varies 
\cite{derT,stauT,hin,sirT}, is a universal exponent for the whole class of models, in
the low temperature region of the plane of parameters.

In section 2 we first focus our interest on the definition of this class of models. 
We then give a qualitative description of its dynamics, showing the existence
of a critical line between a low and a high temperature region in the space of
parameters defining the models (section 3). 
Section 4 is devoted to the study of persistence in the low temperature region.
We first measure the fraction of spins which never flipped up to time $t$, along the
special line joining the voter model to the zero temperature Ising model.
We then discuss the question of persistence at finite temperature (section 4).
Finally an appendix is devoted to the dual description of the models in terms of
reaction diffusion processes.

After this work was completed, we discovered that this class of models
had been previously introduced in \cite{oliv2},
which also contains a determination of the critical line by a finite size
scaling analysis in the stationary state.
For completeness, we kept the original wording of sections 2 and 3,
adding references to this work where appropriate.
%-----------------------------------------------------------------------
\section{Definition of the class of models
\label{model}}
%-----------------------------------------------------------------------
 
Let us consider a two dimensional lattice of spins $\sigma_{i}=\pm 1$, 
evolving with the following dynamical rule. 
At each evolution step, the spin to be updated flips with
the heat bath rule: the probability that the spin $\sigma_i$ takes the value
$+1$ is
$P(\sigma_i=1)=p(h_i)$, where the local field
$h_{i}$ is the sum over neighbouring sites $\sum_{j}\sigma _{j}$ and
\begin{equation}
p(h)=\frac{1}{2}\left(1+\tanh [\beta(h) h]\right)
.\end{equation}
The functions $p(h)$ and $\beta(h)$ are defined over integral values of $h$.
For a square lattice, $h$ takes the values 4, 2, 0, $-2$, $-4$.
We require that
$p(-h)=1-p(h)$, in order to keep the up down symmetry,
hence
$\beta(-h)=\beta(h)$.
Note that this fixes $p(0)=1/2$.
The dynamics therefore depends on two parameters 
\begin{equation}
p_{1}=p(2), \quad p_{2}=p(4)
,\end{equation}
or equivalently on two effective temperatures 
\begin{equation}
\label{defT}
T_1=\frac{1}{\beta(2)}, \quad T_2=\frac{1}{\beta(4)}
.\end{equation}
Defining the coordinate system
\begin{equation}
t_1=\tanh \frac{2}{T_1}, \quad
t_2=\tanh \frac{2}{T_2}
\end{equation}
with $0\le t_1,t_2\le1$, yields
\begin{equation}
p_{1}=\frac{1}{2}\left( 1+t_1\right),
\quad
p_{2}=\frac{1}{2}\left( 1+\frac{2 t_2}{1+t_2^2}\right)
,\end{equation}
with $1/2\le p_1,p_2\le1$.

Hereafter we shall interpret $T_1$ and $T_2$ as two temperatures, respectively
associated to {\it interfacial noise}, and to {\it bulk noise}.
This should be understood in the following sense.
Consider the initial configuration where the system is divided by
a flat interface into two halves, one half with all spins
$\sigma=1$, and the other one with all spins $\sigma=-1$.
Then, if $p_2=p_1=1$, i.e. $T_1=T_2=0$, this configuration will
not evolve in time, neither in the bulk, nor on the interface,
since all spins are surrounded by at least three spins of
the same value. 
However, if $T_2=0$, then as soon as $T_1>0$, spins at the interface will flip,
while those in the bulk will not.
Conversely,  if $T_1=0$, then spins in the bulk will flip if $T_2>0$,
while those on the interface will not.
(They will later do so because of the noise coming from the bulk.)
Note that a configuration where the system is divided into two halves by
a curved interface will always evolve, even if $p_1=p_2=1$, since
$p(0)=\frac{1}{2}$.

Each point in the parameter plane $(p_1,p_2)$, or alternatively in the
temperature plane $(t_1,t_2)$, corresponds to a particular model.
The class of models thus defined comprises as special cases the Ising model, the
voter and antivoter models, as well as the majority vote model,
the description of which follows.
 
The Ising model with heat bath dynamics corresponds to choosing
\begin{equation}
\beta(h)={\rm const.}=\beta,\quad \forall h
,\end{equation}
where $\beta$ is the usual inverse temperature.
Hence
\begin{equation}
p_1=\frac{1}{2}(1+\tanh 2\beta) \quad p_2=\frac{1}{2}(1+\tanh 4\beta) 
.\end{equation}
This corresponds to the line
\begin{equation}
p_2=\frac{p_1^2}{1-2 p_1+2 p_1^2}
\quad\text{or}\quad t_1=t_2
,\label{lineising}\end{equation}
when $\beta$ varies.
For example, at zero temperature, one has $p_1=p_2=1$, hence $T_1=T_2=0$.
The dynamics is therefore only driven by the curvature of the interfaces
between domains of equal values of the spin \cite{brayRevue}.
While the Ising model possesses a well defined energy
for which the heath bath dynamics satisfies
detailed balance,
none of the models outside the Ising line shares these properties \cite{oliv2}, 
hence these models do not possess a usual equilibrium description,
and their definition is purely dynamical.

The dynamics of the voter model is defined as follows \cite{ligg}.
At each time step, the spin to be updated is aligned with one of its neighbours,
chosen at random. 
Therefore 
\begin{equation}
p(h)=\frac{1}{2}(1+\frac{h}{4})
\label{defvot1}
,\end{equation}
i.e.
\begin{equation}
p_1=\frac34,\ p_2=1\quad\text{or}\quad t_1=\frac12,\;t_2=1
.\label{defvot2}
\end{equation}
This definition corresponds to a model with no bulk noise ($p_2=1$, or $T_2=0$).
The noisy voter model is defined as a simple generalization of (\ref{defvot1}). 
The spin to be updated is now aligned with one of its neighbours, chosen at
random, with a probability $\gamma$ \cite{ligg,spohn}. 
In other terms 
\begin{equation}
p(h)=\frac{1}{2}(1+\frac{\gamma}{4}h)
.\label{noisyvoter}
\end{equation}
Hence
\begin{equation}
p_1=\frac{1}{2}(1+\frac{\gamma}{2}) \qquad p_2=\frac{1}{2}(1+\gamma) 
.\end{equation}
This corresponds to the line
\begin{equation}
p_2=2 p_1-\frac12
\quad \text{or}\quad t_1=\frac{t_2}{1+t_2^2}
,\label{linevot}\end{equation}
when $\gamma$ varies from 0 to 1.
One may also extend the definition (\ref{noisyvoter}) to
$-1\le\gamma\le0$, by allowing
negative values of
the coordinates $t_1$ and $t_2$, or equivalently letting $p_1$ and $p_2$ be less
than $1/2$.
The model thus defined is known as the antivoter model \cite{ligg}.

Finally, for the majority vote model \cite{ligg,oliv}, 
spins are aligned with the local field (i.e. with the majority of neighbours)
with some given probability. 
More precisely, if $h\neq 0$
\begin{equation}
p(h)=\frac12\left(1+\delta\, \sign h\right)
\quad (0\le \delta\le1)
,\end{equation}
and $p(0)=\frac{1}{2}$.
Hence
\begin{equation}
p_1=p_2=\frac{1}{2}(1+\delta)
,\end{equation}
therefore the model corresponds to the line
\begin{equation}
p_1=p_2,\quad\text{or}\quad t_1=\frac{2t_2}{1+t_2^2}
,\end{equation}
i.e. $T_2=2\,T_1$, when $\delta$ varies from 0 to 1.

%-----------------------------------------------------------------------
\begin{figure}
\begin{center}
\leavevmode
\epsfxsize=65truemm
\epsfysize=65truemm
\epsfbox{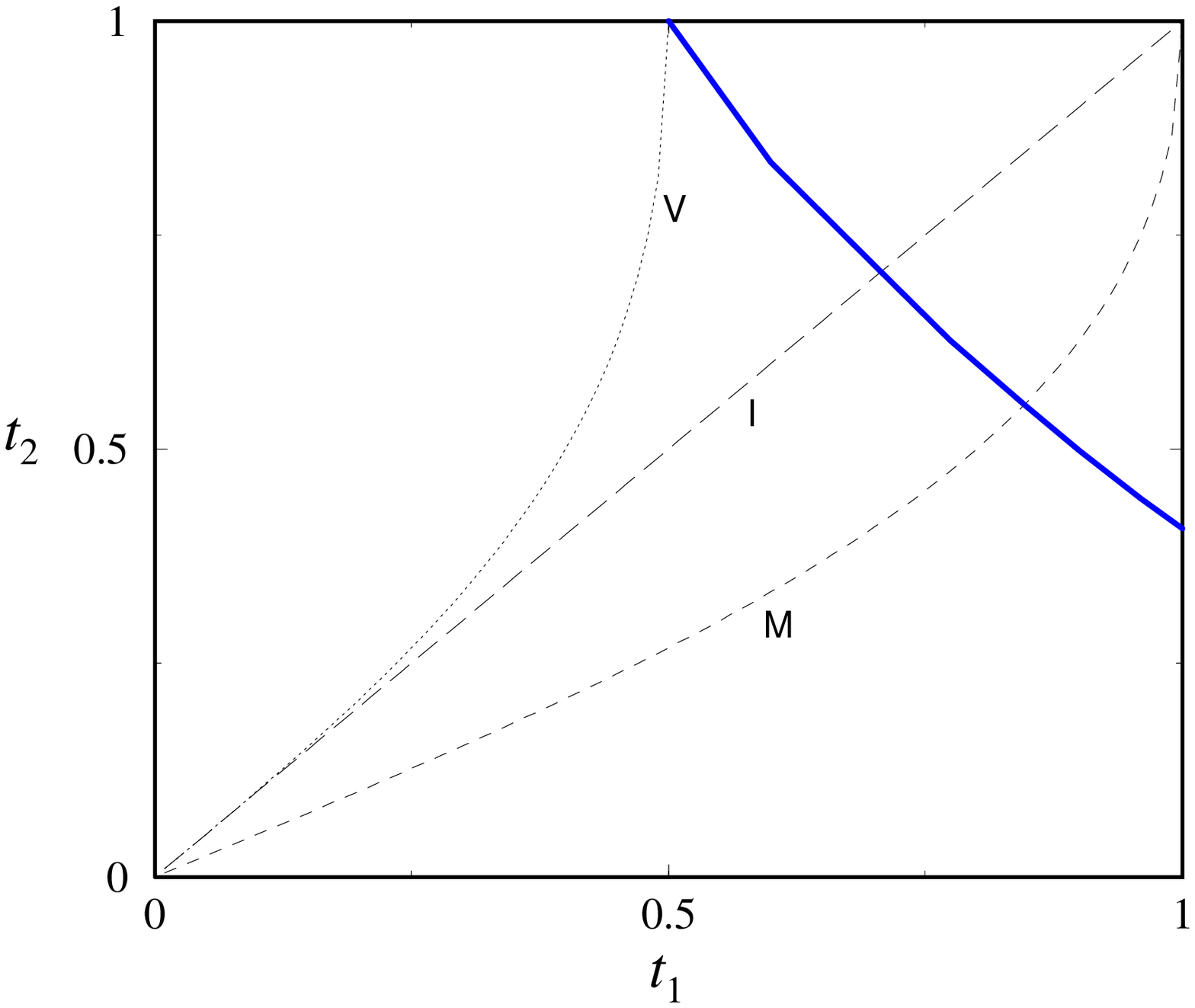}
\epsfxsize=65truemm
\epsfysize=65truemm
\epsfbox{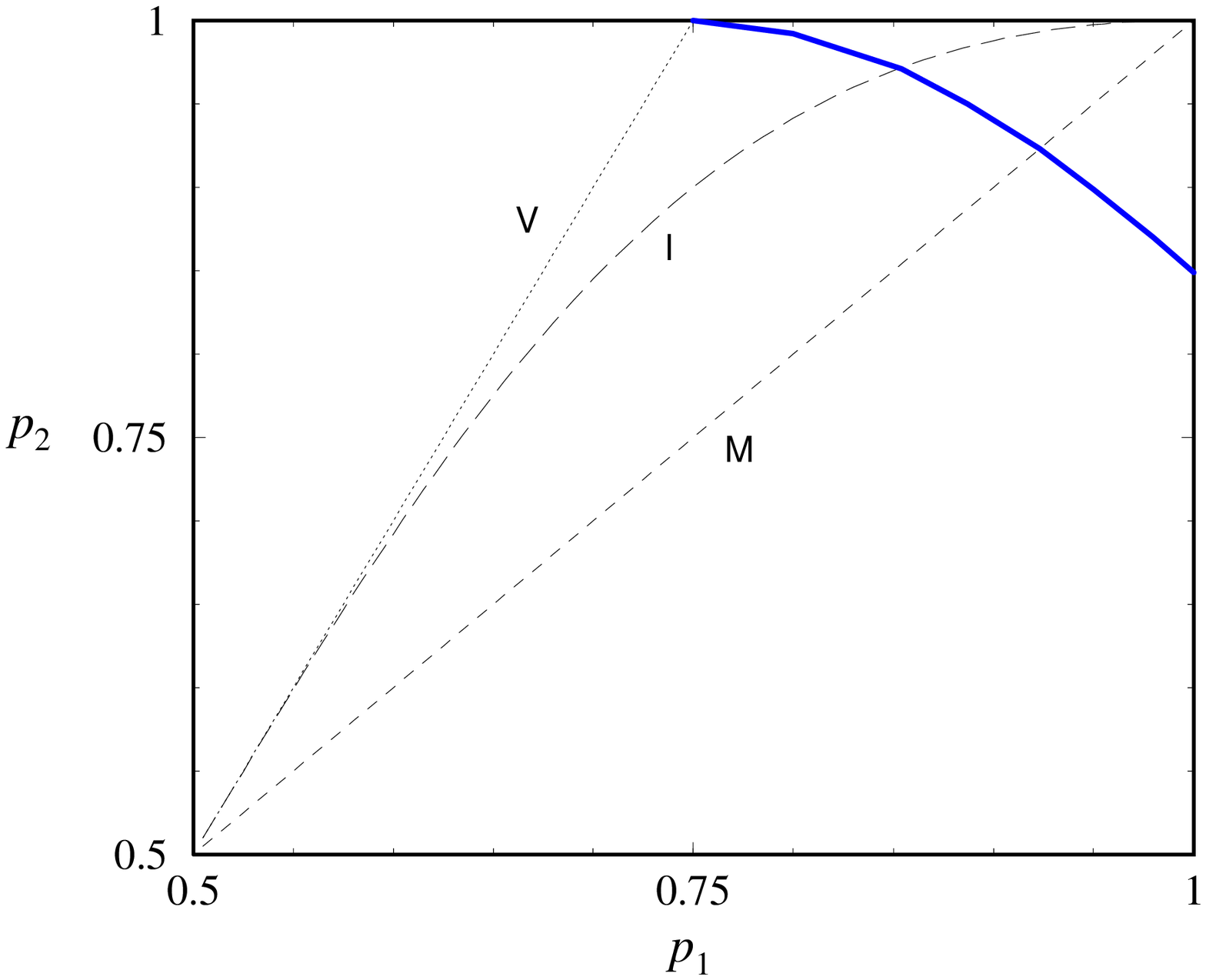}
\end{center}
\caption{Phase diagram.
Broken lines correspond to the noisy voter model (V),
the Ising model (I) and the majority vote model (M). The low temperature
phase is located in the upper right corner, above the transition line (full line).
\label{fig_phase}}
\end{figure}
%-----------------------------------------------------------------------

Figure \ref{fig_phase} shows, in the $(t_1,t_2)$ or $(p_1,p_2$)
planes, the lines corresponding to the Ising, voter and majority vote
models.
Three other lines deserve attention.
Firstly, the $p_2=1$ line corresponds to models with no bulk noise ($T_2=0$),
hence the dynamics is only driven by interfacial noise, defined above.
Secondly, the $p_1=1$ line corresponds to models with no interfacial noise
($T_1=0$), hence the dynamics is only driven by bulk noise.
(In both cases the effect due to the curvature of the interfaces is 
always present, as mentioned above.)
For these last models, the local spin aligns in the
direction of the majority of its neighbours with probability one, 
if the local field $h=2$,
i.e. if there is no consensus amongst the neighbours. 
If there is consensus amongst them, i.e. if $h=4$, the local
spin aligns with its neighbours with a probability $p_2<1$. 
Finally the transition line between the low and high temperature regions is
discussed below.

See reference \cite{oliv2} for similar definitions.
The introduction of the temperatures $T_1$ and $T_2$, together with their interpretations as given above, is original to the present work.
A dual description of the dynamics of the models in terms of reaction diffusion
processes is given in the appendix.

%-----------------------------------------------------------------------
\section{Phase ordering and the critical line}
\label{phase}
%-----------------------------------------------------------------------

In this section we investigate the dynamics of phase ordering for the two parameter
class of models presented here.
For $p_1$ and $p_2$ given, we let the system evolve,
starting from a random initial configuration.
This study demonstrates the existence of a critical line, in the 
$(p_1,p_2)$ or $(t_1,t_2)$ planes,
between a low temperature region where clusters (or domains) grow indefinitely,
and a high temperature region where one only observes fluctuations at a
finite scale. 
It also provides a visual illustration of the role of the two temperatures $T_1$ and $T_2$, respectively associated to interfacial and to bulk noise.

Two methods for updating the spins are at our disposal.
With sequential updating, time is continuous and each spin evolve
independently of others, at times chosen randomly according to a Poissonian
law. 
The normalization is chosen such that each spin evolves once in a mean
unit of time. 
Therefore, in a simulation of a lattice of $N$ spins, one has to choose randomly
a spin and update it, $N$ times per time unit.

Instead of using sequential updating, 
we adopt a slightly different procedure for our simulations,
namely a parallel updating of the spins. 
The lattice is divided into two odd and even sublattices, and, during a
unit of time, the two sublattices are visited in turn, with a systematic update
of their spins.
This division is needed in order to avoid undesirable effects due to the
simultaneous update of neighbouring sites. 
This process allows a functional parallelization on the computer. 
Spins are represented as single bits, and
it is possible to arrange the spins into computer words, such that all spins
of a word can be updated simultaneously by global logical operations. 
This method is numerically efficient, allowing for long simulations on large
lattices.

In figures 2--7, we have selected snapshots of the same fragment $256\times 256$
of a $512\times 512$ lattice with periodic boundary conditions, for three times
$t=8$, 64 and 512, and six different points in the parameter space of the models.

\begin{figure}
\begin{center}
\leavevmode
\epsfxsize=\hsize
\epsfbox{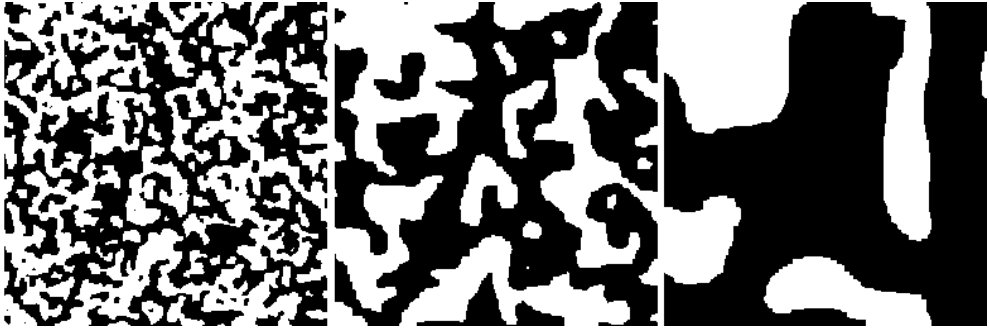}
\end{center}
\caption{
Snapshots of the evolution for the Ising model at zero temperature ($p_1=p_2=1$).
\label{evi0}}
\end{figure}

\begin{figure}
\begin{center}
\leavevmode
\epsfxsize=\hsize
\epsfbox{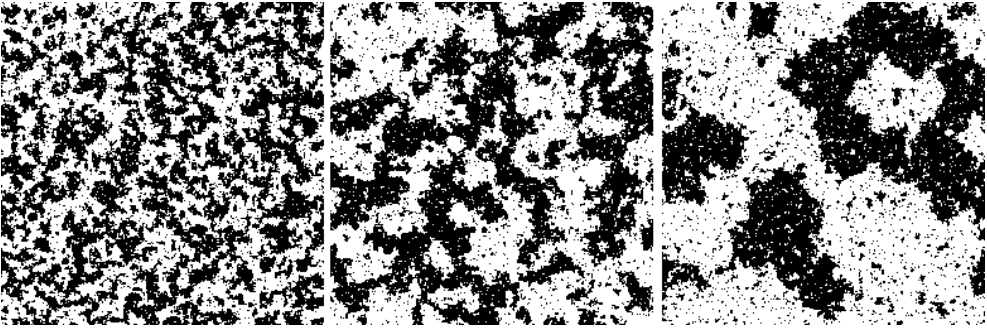}
\end{center}
\caption{
Snapshots of the evolution for the Ising model at $T=0.97\,T_c$ 
($p_1=0.8602$, $p_2=0.9743$).
\label{evi97}}
\end{figure}

Figures \ref{evi0}, \ref{evi97} depict, as reference views, the
well known appearance of domain growth for the Ising model, respectively at zero
temperature, and at
$T=0.97\,T_c$.
The Ising critical point corresponds to $t_1=t_2=1/\sqrt{2}$.
For $T<T_c$ clusters grow as $\sqrt{t}$, while at $T_c$, they only grow as
$t^{1/z}$, where the dynamical exponent $z\approx 2.17$.
In the high temperature phase, clusters grow until their sizes reach the
equilibrium correlation length. 

%-----------------------------------------------------------------------
%-----------------------------------------------------------------------

%-----------------------------------------------------------------------
%-----------------------------------------------------------------------

This distinction between a low temperature and a high temperature phase can be
generalized to all models of the two dimensional parameter plane.
The transition line between the low and high temperature regions, depicted in
figure \ref{fig_phase}, was located by a systematic investigation of the
parameter plane. 

\begin{figure}
\begin{center}
\leavevmode
\epsfxsize=\hsize
\epsfbox{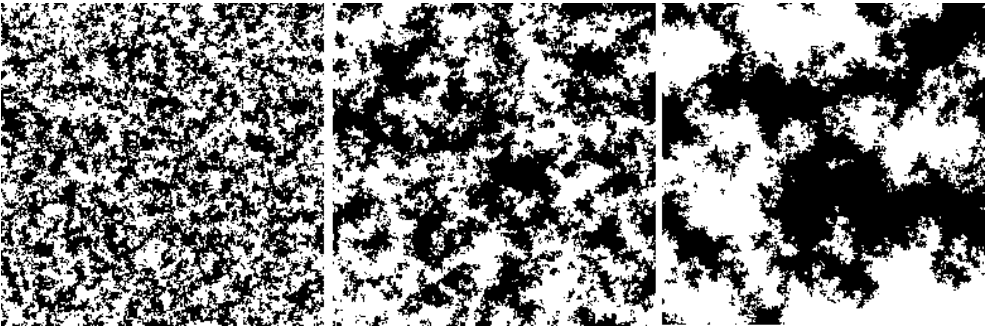}
\end{center}
\caption{
Snapshots of the evolution for a model close to the
voter model, in the low temperature region
($p_1=0.76$, $p_2=1$).
\label{evv76}}
\end{figure}

\begin{figure}
\begin{center}
\leavevmode
\epsfxsize=\hsize
\epsfbox{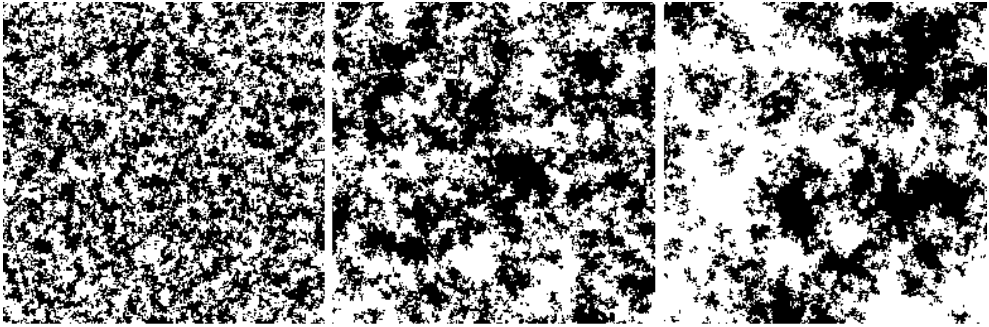}
\end{center}
\caption{
Snapshots of the evolution for the voter model ($p_1=0.75$, $p_2=1$).
\label{evv75}}
\end{figure}

\begin{figure}
\begin{center}
\leavevmode
\epsfxsize=\hsize
\epsfbox{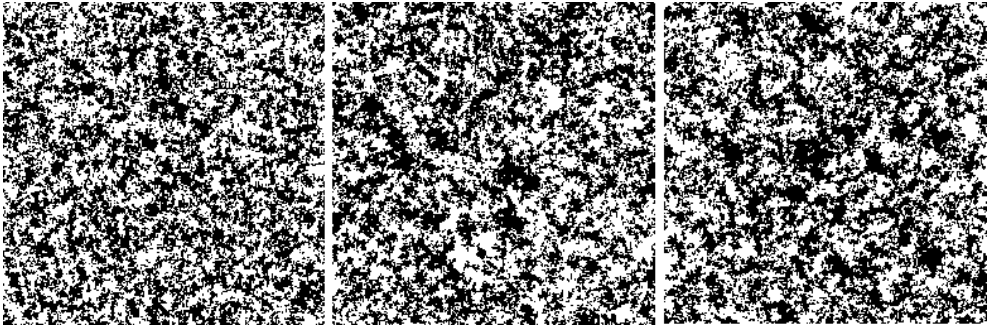}
\end{center}
\caption{
Snapshots of the evolution for a model close to the voter model, in the high
temperature region
($p_1=0.72$, $p_2=1$).
\label{evv72}}
\end{figure}

For instance, figures \ref{evv76}, \ref{evv75}, and \ref{evv72} correspond to an
exploration of the line
$p_2=1$, with $p_1$ decreasing away from the zero temperature Ising case
($p_1=p_2=1$). 
On this line no thermal fluctuations occur inside the clusters, so their
interiors remain forever uniformly black or white. 
When $p_1$ decreases, the boundaries between domains take progressively
a fractal appearance, in accord with the fact that interfacial noise
increases along the line.
This is particularly visible on figure \ref{evv76}, which illustrates the case
of a model close to the voter model, in the low temperature region ($p_1=0.76$,
$p_2=1$), and on figure \ref{evv75}, for the voter model itself
($p_1=0.75$, $p_2=1$).
Again one observes growing clusters, with a characteristic size behaving as
$\sqrt{t}$.
A similar behaviour is observed for all models such that
$0.75\le p_1\le 1, p_2=1$. 

%-----------------------------------------------------------------------
%-----------------------------------------------------------------------

%-----------------------------------------------------------------------
%-----------------------------------------------------------------------

Figure \ref{evv72} illustrates the case of a model located in the high
temperature region, near the voter model ($p_1=0.72$, $p_2=1$).
In parallel to what is observed for the Ising case above $T_c$,
clusters grow until their sizes reach the correlation length. 
The transition to the low temperature region along the line $p_2=1$ is
therefore only induced by the interfacial noise.

%-----------------------------------------------------------------------
%-----------------------------------------------------------------------

\begin{figure}
\begin{center}
\leavevmode
\epsfxsize=\hsize
\epsfbox{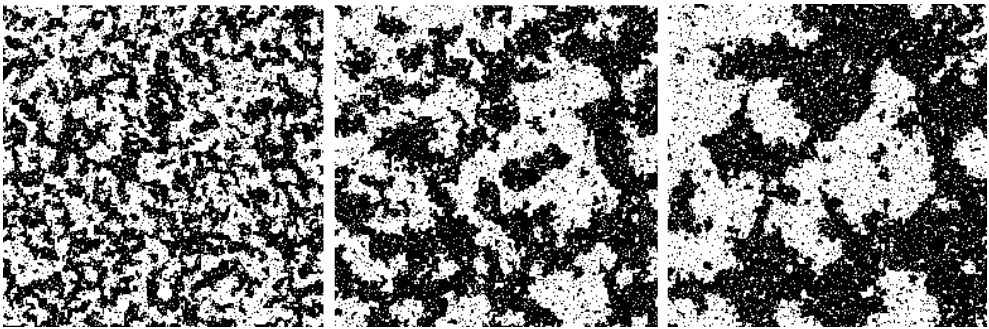}
\end{center}
\caption{
Snapshots of the evolution for the majority vote model ($p_1=p_2=0.93$).
\label{evm93}}
\end{figure}

We also display in figure \ref{evm93} the evolution observed
for the majority vote model with ($p_1=p_2=0.93$).
The transition takes place for $p_1=p_2\approx 0.923$, in agreement with the
value ($0.925$) given in \cite{oliv}.

%-----------------------------------------------------------------------
%-----------------------------------------------------------------------

\begin{figure}
\begin{center}
\leavevmode
\epsfxsize=\hsize
\epsfbox{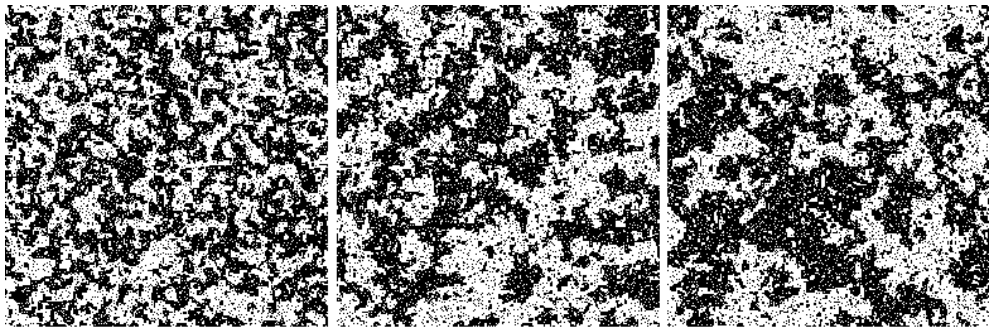}
\end{center}
\caption{
Snapshots of the evolution for the model with ($p_1=1$, $p_2=0.84$).
\label{evr84}}
\end{figure}

Finally exploring the line $p_1=1$, we observe a transition between the low and high temperature regimes  for $p_2\approx 0.849$, 
in agreement with the value ($0.854$) given in \cite{oliv2}.
Figure \ref{evr84} depicts snapshots of the evolution for the model with
($p_1=1$, $p_2=0.84$).
See \cite{oliv2} for a determination of the critical line by a finite size scaling analysis in the stationary state.

%-----------------------------------------------------------------------
%-----------------------------------------------------------------------

%-----------------------------------------------------------------------
\section{Persistence
\label{persist}}
%-----------------------------------------------------------------------

In this section we address the question of persistence in the low temperature
region of the ($p_1,p_2$) parameter plane depicted in
figure \ref{fig_phase}. 
We first present numerical measurements of the fraction $R(t)$ of spins which never
flipped up to time $t$, for models with no bulk noise, i.e. for $T_2=0$ or $p_2=1$.
The case of persistence at finite temperature, i.e. in the present context for $T_2>0$,
will be discussed in section \ref{perT}.

\subsection{The $p_2=1$ line}
\label{p21}

\begin{figure}
\begin{center}
\leavevmode
\epsfysize=80truemm
\epsfbox{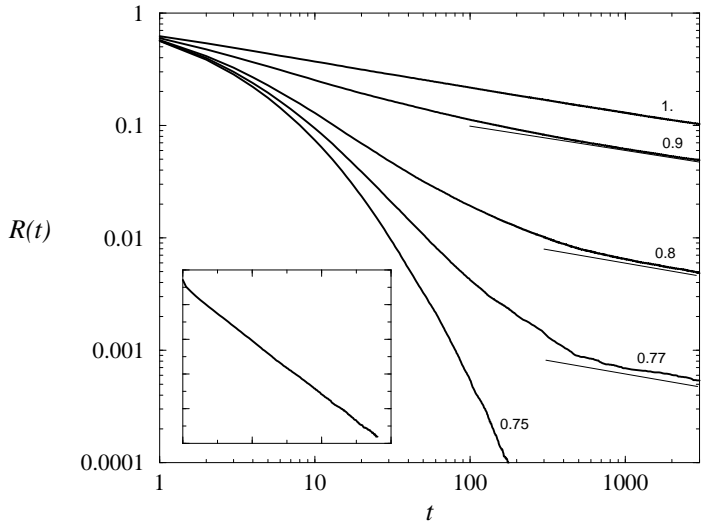}
\end{center}
\caption{Log-log plot of the persistence probability $R(t)$ on the line $p_{2}=1$, for
$p_1=1,0.9, 0.8, 0.77, 0.75$, from top to bottom.
The inset depicts $\ln R(t)$ for the voter model ($p_1=0.75$) against 
$\ln^2 t$.
The system size is $L=3072$.
\label{perss}}
\end{figure}

We measured persistence 
in the low temperature region ($p_1\ge 0.75$) of the line $p_2=1$.
Figure \ref{perss} depicts the results of
simulations for times up to 5000, on a lattice $3072\times 3072$.
The fraction $R(t)$ of spins which never flipped up to time $t$ behaves, in the early
stage, as 
\begin{equation}
\ln R(t)\sim -A \ln^2 t+B\ln t+C+\cdots 
\label{perslaw1}
,\end{equation}
then one observes a crossover toward the power law 
\begin{equation}
R(t)\sim t^{-\theta } 
\label{perslaw2}
,\end{equation}
with an exponent $\theta\approx 0.22 $ which seems independent of $p_{1}$. 
This behaviour therefore interpolates between the two `pure' cases, namely the zero
temperature Ising model ($p_1=p_2=1$) where (\ref{perslaw2}) is observed
with the same exponent $\theta\approx 0.22 $ \cite{der1,stau1,der2}, 
and the voter model, for which (\ref{perslaw1}) holds \cite{ben,how}. 

%-----------------------------------------------------------------------
%-----------------------------------------------------------------------
 
The origin of these two `pure' behaviours 
can be simply traced back to the two driving forces of the dynamics
in absence of bulk noise, i.e. respectively to interfacial noise for (\ref{perslaw1}),
and to the curvature of interfaces for (\ref{perslaw2}).
Indeed, consider first the case of the Ising model at zero temperature ($T_1=T_2=0$),
for which the dynamics is only driven by the curvature of interfaces.
As a consequence, the only flipping mechanism for a spin is the slow sweeping mode of
the interfaces, hence persistent spins deeply buried inside large clusters disappear
slowly, according to (\ref{perslaw2}).
Consider now the case of the voter model ($p_1=0.75, p_2=1$), starting from a
configuration with a planar interface separating two regions of opposite spins. 
This interface thickens gradually, because of interfacial noise, taking ever more a
fractal appearance. 
At large times it becomes difficult to speak any longer of an
interface, the thickness of which should be considered as infinite.
The decay of persistent spins is much faster and follows (\ref{perslaw1}).

This analysis allows the following interpretation of the results given above 
for persistence of a generic model along the
$p_2=1$ line, and $p_1>0.75$. 
In the early stages of coarsening, domains begin to grow
and develop their interfaces. 
As a consequence, the fraction of persistent spins in the interface region is expected
to decay more rapidly than if they were deep inside a cluster,
following (\ref{perslaw1})
since this is the dominant process.
The duration of this early stage is related to the thickness of the interfaces,
which vanishes for the Ising model where smooth interfaces are observed, and diverges
for the voter model where a fractal appearance of the interfaces is observed. 
After this first stage, the fraction of persistent spins will decay 
according to (\ref{perslaw2}). 
Therefore, along the line $p_2=1$ ($T_2=0$), the resulting law for the decay of $R(t)$
will be a weighted sum of (\ref{perslaw1}) and (\ref{perslaw2}). 
The amplitude obtained by fitting the asymptotic algebraic tail (\ref{perslaw2}) is
naturally interpreted as the fraction of persistent spins which are deep inside the
clusters. 
This fraction vanishes when one approaches the voter model, with an apparent
power law $(p_{1}-0.75)^{\alpha}$, with $\alpha\approx 2.3$. 
This result is consistent with the fact that the voter model is critical, according to
the criterion of section \ref{phase}.

Note that, for any model such that $p_1>0.75$, the prescription of suppressing the bulk noise by forbidding flips of a spin $\sigma=1$ (resp. $\sigma=-1$)
surrounded by four spins with the same value $1$ (resp. $-1$),
leads to an algebraic decay of the persistence probability.
This corresponds to a projection of the model onto the line $p_1=1$.

\subsection{Persistence at finite temperature}
\label{perT}

For a two dimensional Ising system in presence of thermal noise 
(i.e. in the present context, for $p_2<1$),
the number of spins which did not flip up to time $t$ decays exponentially. 
A more satisfactory quantity to investigate is the fraction
$R(t)$ of space which remained in the same phase, coming back to the
original definition of persistence \cite{marcos,bray},
where `remaining in the same phase' means remaining `dry', i.e. unswept by an
interface. 
The prescription given in \cite{derT} to determine whether a given spin
`remained in the same phase', leads again to
algebraic decay of $R(t)$, the fraction of `persistent' spins, when $T<T_c$.
Further works \cite{stauT,hin,sirT} seem to indicate that the
persistence exponent $\theta$ for the two dimensional Ising model is constant in the low
temperature phase, and approximately equal to 0.22.

%-----------------------------------------------------------------------
\begin{figure}
\begin{center}
\leavevmode
\epsfysize=80truemm
\epsfbox{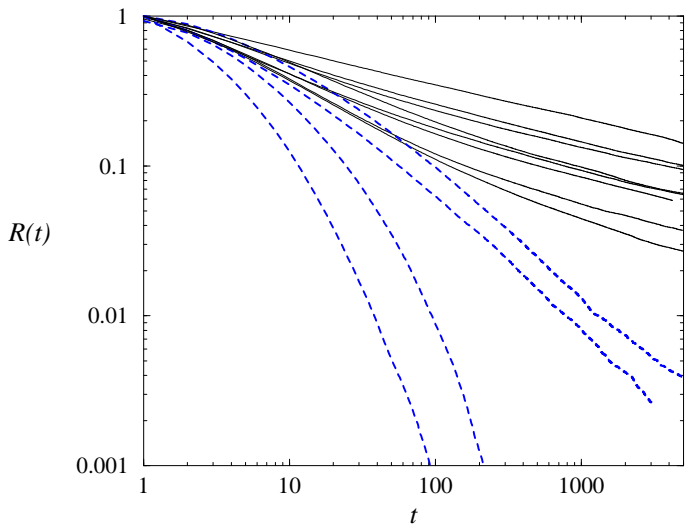}
\end{center}
\caption{Log-log plot of the persistence probability $R(t)$.
Full lines, from top to bottom:
($p_1=1, p_2=1$): Ising at $T=0$ ;
($p_1=0.95, p_2=0.997$): Ising at $T=0.6\, T_c$;
($p_1=1, p_2=0.95$);
($p_1=0.95, p_2=0.95$): Majority vote;
($p_1=0.9, p_2=1$);
($p_1=0.8, p_2=0.988$): Ising at $T=0.8\, T_c$;
($p_1=1, p_2=0.9$);
($p_1=0.876, p_2=0.980$): Ising at $T=0.9\, T_c$;
Dashed lines, from top to bottom:
($p_1=0.92, p_2=0.92$): Majority vote;
($p_1=0.854, p_2=0.971$): Ising at $T=T_c$;
($p_1=1, p_2=0.84$);
($p_1=0.75, p_2=1$): Voter.
The system size is $L=1536$.
\label{tous}}
\end{figure}
%-----------------------------------------------------------------------

Coming back to the class of models under investigation, 
one may therefore wonder what is the value of the persistence exponent
outside the Ising line, in the low
temperature region. 
Here we use the definition of \cite{derT}, adopting the extensions
given in
\cite{hin}, in order to measure persistence in the low temperature region of the
parameter plane. 
We consider three copies of the system, A, B and C, submitted to the
same noise ($p_1,p_2$).
The initial condition for A is random, while it is ordered for B 
($\sigma_i^{\rm B}(0)=+1$), and C ($\sigma_i^{\rm C}(0)=-1$) where $i=1,\ldots, N$
labels the sites.
A spin $\sigma_i^{\rm A}$ of copy A is said to be persistent up to time $t$, if it
experienced thermal fluctuations only, i.e. if its history was either that of 
$\sigma_i^{\rm B}$ or that of $\sigma_i^{\rm C}$.
Hence 
\begin{equation}
R(t)=P\big(\sigma_i^{\rm A}(t')=\sigma_i^{\rm B}(t'),\ \text{or}\
\sigma_i^{\rm A}(t')=\sigma_i^{\rm C}(t'),\forall t'\le t\big)
\label{rtnew}
\end{equation}
measures the fraction of persistent spins in one of the two phases $\pm$
\cite{hin,derT}.
We also checked, using the determination of the interfaces proposed in \cite{hin}, 
the equivalence between the definition of $R(t)$ given by (\ref{rtnew}) and that
obtained by considering a spin as persistent if it was not swept by an interface. 

Figure \ref{tous} depicts the results.
The exponent $\theta$ seems constant, with a value $\approx 0.22$, for 
all models on the low temperature side of the critical line that we considered.
It is nevertheless hard to conclude, on the basis of numerical measurements. 
The decrease of persistence at criticality is much faster.
A possible explanation is that, as for the voter model (see \ref{p21}), the thickness of
interfaces diverges, leading to a behaviour similar to (\ref{perslaw1}).

%-----------------------------------------------------------------------
\section{Conclusion
\label{conclusion}}
%-----------------------------------------------------------------------
The aim of the present work was to initiate the study of coarsening and persistence
for a class of models interpolating between the voter and Ising models,
introduced previously in \cite{oliv2}.

In our opinion, the introduction of two temperatures $T_1$ and $T_2$, respectively
associated to interfacial and bulk noise, gives a useful interpretation of the phenomena
observed when varying continuously the parameters defining this class of models.
The present analysis seems to indicate that
$\theta\approx 0.22$ is the universal persistence exponent for all models in the low
temperature phase. 
The situation on the critical line is less clear, and possibly
related to the divergence of the interfaces between domains at criticality.
In this respect it would be desirable to find a way of computing the equation of
the critical line in the $(p_1,p_2)$ parameter plane.
A more ambitious goal would be to compute $\theta$ for the two
dimensional zero temperature Ising model, in the dual reaction diffusion framework given
in the appendix, using the field theoretical methods of \cite{lee,howard,cardy,how}. 
Let us finally mention extensions to $d>2$ or to the Potts model as possible further
studies. 
We hope to address some of those points in the future.

\ack
We wish to thank M Howard and J-M Luck for interesting discussions.

%-----------------------------------------------------------------------
\appendix
\section{The dual approach}
%-----------------------------------------------------------------------

In this appendix we show that the dynamics of this class of models 
possesses a dual description in terms of reaction diffusion processes
for a set of random walkers moving backward in time.
This description, beyond its intuitive value,
is the starting point of an analysis of the temporal evolution of the system by means
of diagrammatic expansions
\cite{unpub}.

We first review the case of the voter model \cite{ligg,cox,spohn,der3}.
Let us define the random variable $\tau=(1+\sigma)/2=1,0$, hereafter
also called spin.
The value of the spin $\tau^{(t)}_{x}$ at time $t$ and site $x$ can be traced back
to the value of its ancestor $\tau_{y}^{(0)}$ at time $t=0$, located on site
$y$.
The idea is to follow the line of continuity of the value of the spin, backward
in time, from ($x, t$) to ($y,0$).
Indeed, one first follows the line of constant $\tau$
backward in time, from
$(x,t)$ until the last updating of site $x$ is met. 
At this time, the spin had chosen the value of a neighbouring site.
Moving to this site one again follows the line of constant $\tau$ backward in
time. 
Proceeding until $t=0$, one thus performs a random walk, backward in time, from
($x,t$) to ($y, 0$) such that $\tau_{x}^{(t)}=\tau_{y}^{(0)}$.

It is easy to see that all spins at time $t$ possess an ancestor at time $t=0$,
but that conversely not all sites at time $t=0$ are ancestors of spins at
time $t$.
Indeed, starting from $n$ spins 
$\tau^{(t)}_{1},\ldots, \tau^{(t)}_{n}$, located on sites $x_1,\ldots, x_n$, at
time $t$, the same reasoning leads to the consideration of $n$ random walkers
starting from sites $x_1,\ldots, x_n$ at time $t$, and going backward in time.
When two such walkers meet, they coalesce because the corresponding
sites have a common ancestor from which they inherit the common value of
their spin. 

%-----------------------------------------------------------------------
\begin{figure}
\begin{center}
\leavevmode
\epsfysize=50truemm
\epsfbox{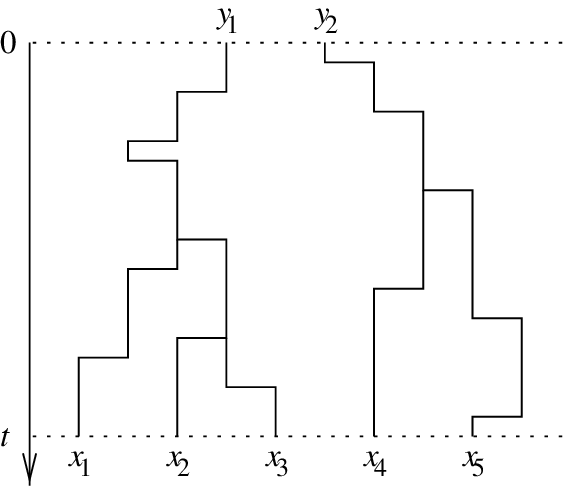}
\end{center}
\caption{An example of coalescing random walks moving backward in time, occurring
in the voter model.
The drawing is done in one dimension, for simplicity.
\label{fig5}}
\end{figure}
%-----------------------------------------------------------------------

Hence the determination of the values of these $n$ spins is equivalent, for the
voter model, to knowing the history of $n$ independent coalescing random walkers,
starting from sites $x_1,\ldots,x_n$, and moving backward in time.
An example is displayed in figure \ref{fig5}. 
The final stage, at time $t=0$, contains $k$ walkers
($1\leq k\leq n$) at sites 
$y_1,\ldots,y_k$. 
Since 
$\tau_{x_{1}}^{(t)}\cdots \tau_{x_{n}}^{(t)}=\tau _{y_{1}}^{(0)}\cdots
\tau _{y_{k}}^{(0)}$, 
we conclude that the sum of the probabilities of all
these processes starting from the $n$ points $x_{1},\dots,x_n$
gives the correlation function
$\langle\tau _{x_{1}}^{(t)}\cdots \tau_{x_{n}}^{(t)}\rangle$.
One can easily generalize the reasoning to the case of spins on
different sites $x_1,\ldots, x_n$, at different times $t_1,\ldots, t_n$.

For the voter model, the definition (\ref{defvot1})
can be rewritten as
\begin{equation}
P(\tau _{x}=1)=\frac{1}{4}\left( \tau _{y_{1}}+\tau _{y_{2}}+\tau
_{y_{3}}+\tau _{y_{4}}\right) 
.\label{vdef}
\end{equation}
This means that for a two dimensional random walk the elementary process
consisting of a jump from site $x$ to one of its neighbouring sites
$y$ has probability $\frac{1}{4}\tau_{y}$.
Similarly (\ref{noisyvoter}) leads to 
\begin{equation}
P(\tau _{x}=1)=
1-p_{2}
+(p_{2}-p_{1})\left(\tau_{y_1}+\tau_{y_2}+\tau_{y_3}+\tau_{y_4}\right)
\label{vvdef}
,\end{equation}
the interpretation of which is given below.
The generalization to a generic model ($p_1,p_2$) is now straightforward. 
By its very definition one has 
\begin{equation}
\begin{array}{l}
P(\tau _{x}=1)= p_{2}\tau _{y_{1}}\tau _{y_{2}}\tau _{y_{3}}\tau _{y_{4}}\\ 
\quad +p_{1}\left[ \tau _{y_{1}}\tau _{y_{2}}\tau _{y_{3}}\left( 1-\tau
_{y_{4}}\right) +\text{3 perm. terms}\right] \\ 
\quad +\frac{1}{2}\left[ \tau _{y_{1}}\tau _{y_{2}}\left( 1-\tau _{y_{3}}\right)
\left( 1-\tau _{y_{4}}\right) +\text{5 perm. terms}\right] \\ 
\quad +(1-p_{1})\left[ \tau _{y_{1}}\left( 1-\tau _{y_{2}}\right) \left( 1-\tau
_{y_{3}}\right) \left( 1-\tau _{y_{4}}\right) +\text{3 perm. terms}\right]
\\ 
\quad +(1-p_{2})\left( 1-\tau _{y_{1}}\right) \left( 1-\tau _{y_{2}}\right)
\left( 1-\tau _{y_{3}}\right) \left( 1-\tau _{y_{4}}\right)
\end{array}
\end{equation}
This expression can be rewritten as 
\begin{equation}
\begin{array}{l}
P(\tau _{x}=1)=1-p_{2} \\ 
\qquad +(p_{2}-p_{1})\left( \tau _{y_{1}}+\tau _{y_{2}}+\tau _{y_{3}}+\tau
_{y_{4}}\right) \\ 
\qquad +\left( 2p_{1}-p_{2}-\frac{1}{2}\right) \left(
\tau _{y_{1}}\tau_{y_{2}}+\tau _{y_{1}}\tau_{y_3}+\tau_{y_1}\tau _{y_4}
+\right.\\
\qquad\qquad\qquad\left.+
\tau_{y_2}\tau _{y_3}+\tau _{y_2}\tau _{y_4}+\tau _{y_{3}}\tau_{y_4}\right) \\ 
\quad -\left( 2p_{1}-p_{2}-\frac{1}{2}\right)
\left( \tau_{y_1}\tau_{y_2}\tau_{y_3}+
\tau_{y_1}\tau_{y_2}\tau_{y_4}+\tau_{y_1}\tau_{y_3}\tau_{y_4}+
\tau _{y_2}\tau _{y_3}\tau_{y_4}\right)
\end{array}
\label{mdef}
\end{equation}
Equation (\ref{mdef}) is the generalization of (\ref{vvdef}) and can be
expressed as the sum of four elementary processes, 
namely:

\begin{itemize}
\item the random walker disappears (with a weight $1-p_{2}$),

\item the random walker makes a step to a neighbouring site (with a weight 
$p_{2}-p_{1}$),

\item the random walker splits into two walkers located on different
neighbouring sites (with a weight $2p_{1}-p_{2}-\frac{1}{2}$),

\item the random walker splits into three walkers located on different
neighbouring sites (with a weight $-2p_{1}+p_{2}+\frac{1}{2}$).
\end{itemize}

Hence the coalescing random walks considered above for the voter model
have now also the
possibility of annihilating, or of branching into two or three new walks. 
Note that the weight $p_2-p_1$ vanishes on the line of the majority vote
model, while the weight $2p_1-p_2-\frac{1}{2}$ vanishes on the line of the noisy
voter model.

Let us underline the fact that the weights corresponding to these elementary processes
can be negative. 
Therefore these drawings should be seen as diagrams, and not as the
actual paths of real random walkers. 
They may indeed be used for actual
diagrammatic computations. 
For example, near the voter model, one
can perform a double series expansion of the correlation functions in the
`coupling constants' 
$1-p_{2}$ and $2p_{1}-p_{2}-\frac{1}{2}$, in order to compute the rate of the
exponential decay of persistence, in good agreement with numerical measurements 
\cite{unpub}.

%-----------------------------------------------------------------------

\end{document}